\documentstyle[prd,aps]{revtex}

\def\ltap{\raisebox{-.55ex}{\rlap{$\sim$}} \raisebox{.4ex}{$<$}}
\def\gtap{\raisebox{-.55ex}{\rlap{$\sim$}} \raisebox{.4ex}{$>$}}
\def\gsim{\mathrel{\gtap}}
\def\lsim{\mathrel{\ltap}}

\begin{document}
%\draft
\input epsf

\twocolumn[\hsize\textwidth\columnwidth\hsize\csname
@twocolumnfalse\endcsname

\title{Photons as Ultra High Energy Cosmic Rays ?}
\author{
O.E.~Kalashev$^{(1)}$, V.A.~Kuzmin$^{(1)}$, D.V.~Semikoz$^{(1,2)}$, 
I.I.~Tkachev$^{(1,3)}$\\
$^{(1)}${\small\it Institute for Nuclear Research of the 
Academy of Sciences of Russia,} \\ {\small\it Moscow 117312, Russia}\\
$^{(2)}${\small\it Max-Planck-Institut f\"ur Physik
(Werner-Heisenberg-Institut),}\\{\small\it F\"ohringer Ring 6,
80805 M\"unchen, Germany}\\
 $^{(3)}${\small\it  CERN Theory  Division, CH-1211 Geneva 23, Switzerland.}
}
%\date{}
\maketitle

\begin{abstract}
We study spectra of the Ultra High Energy Cosmic Rays 
assuming primaries are protons and photons, and that their 
sources are extragalactic. 
We assume power low for
the injection spectra and take into account the influence of
cosmic microwave, infrared, optical and radio backgrounds as well 
as extragalactic 
magnetic fields on propagation of primaries. Our additional free parameters
are the maximum energy of injected particles and the distance to the nearest
source. We find a parameter range where the 
Greisen-Zatsepin-Kuzmin cut-off is avoided.
\end{abstract}

\vskip2pc]

\paragraph*{Introduction.} 

Measurements of the spectra of the Ultra High Energy Cosmic Rays (UHECR) 
\cite{spectra} show that
the Greisen-Zatsepin-Kuzmin (GZK) cutoff \cite{GZK} is absent. 
The resolution of the arising puzzle seems to be impossible
without invoking new physics or extreme astrophysics, for reviews
see Refs. \cite{reviews1,reviews2}. Clearly, the GZK cutoff is avoided if the 
distribution of sources is peaked in our
local cosmological neighborhood, or if the primary particles are immune to 
the Cosmic Microwave Background Radiation (CMBR). First possibility
can be realized e.g. in the model of decaying superheavy dark matter 
clustered in the galaxy halo \cite{SHDM}. Second possibility requires
either new hypothetical particle \cite{messengers}, or 
violation of the Lorentz
invariance \cite{VLI}, or extreme neutrino luminosity
if Z-burst model is employed \cite{neutrinos}. 

In this Letter we address the question: is it really impossible to avoid the 
GZK-cutoff within frameworks of the standard physics if
the distribution of (astrophysical) sources is homogeneous ? 
At a first glance, the answer is negative. However,
one should be more careful and recall that in the models of decaying
topological defects the sources are homogeneously distributed throughout the
Universe while the products of the decay are all standard particles.
Nevertheless, the model proves to be working \cite{TD}.
Prime reasons for success are specific injection spectra and a large
fraction of UHE photons.
Can the same work for astrophysical sources ? One can argue
against this conjecture
that topological defects are invisible, except of their UHECR
flux, and therefore it is possible that the distance to the closest
fraction of the decaying defect network is not large. 
This fraction may make a dominant contribution
even if the network itself is homogeneous.
In contrast to this, there are no suitable astrophysical sources within
the GZK sphere. This argument is valid, but not without a caveat. 
Astrophysical sources
may be ``invisible'' as well. Among suggested candidates are gamma-ray bursts 
\cite{GRB} and ``dead'' quasars \cite{DQ}. Both form a homogeneous population
and were assumed to work without invoking new physics. Prime motivation
was invisibility of these sources which resolves the puzzle that rays
do not point back to any visible candidate sources within the GZK sphere.
However, because of the homogeneous distribution 
of respective sources these models should exhibit the GZK cut-off in general.
The analysis of whether it is possible to avoid the cut-off and under 
which conditions was not carried out. The same danger exists in Z-burst models
as well. Indeed, Z-bursts occur homogeneously throughout the Universe
(unless background neutrino are extremely clumped)
which provides a homogeneous source of protons and photons and therefore
the model is subject to the GZK cut-off in principle. 

Recently, highly significant correlations of arrival directions of the UHECR 
with BL Lacertae were found \cite{TT_BL}. Distances to more than a half of
BL Lacs are not known. Closest BL Lacertae with known redshifts are
at $z \sim 0.03$ and therefore outside of the GZK sphere. 
Do such correlations require new physics ? To be sure one should
first firmly exclude standard model particles without making strong
assumptions on the injection spectra.

\paragraph*{Methods.} 

We use numerical code which was
developed in \cite{kks1999}. We calculate 
propagation of protons and photons using standard dominant processes 
(for details see \cite{reviews1}).  
For protons we took into account single and multiple pion production,
and $e^{\pm}$ pair creation. For photons we considered $e^{\pm}$ pair 
production, inverse Compton scattering and double pair production processes.
For electrons and positrons we took into account Compton scattering, 
triple pair production and synchrotron energy loss on EGMF.
Propagation of 
protons and photons is calculated self-consistently. Namely,
secondary (and higher generation) particles arising in all 
reactions are propagated alongside with the primaries.
UHE protons and photons lose their energy in interactions with 
the electro-magnetic background,
which consist of CMBR, radio, infra-red and optical components,
as well as Extra Galactic Magnetic Fields (EGMF).
Protons are sensitive essentially to CMBR only, while
for photons all components of the electro-magnetic background are important. 
We take a minimal model for the radio background
\cite{RB}. For calculating the infra-red/optical background
we used the same approach as in \cite{Lee}.
For the extragalactic magnetic field only the upper bound is 
established observationally, $B < 10^{-9} {\rm G} (l_c/{\rm Mpc})^{1/2}$ \cite{FR}. 
It is believed that galactic magnetic fields can be generated 
from the extragalactic ``seed'' if  
the later has magnitude in the range $B = 10^{-12} - 10^{-9} G$, 
but in some regions it can be much smaller (voids) or larger 
(sheets). In our simulations we vary magnetic field
strength in the range $B = 10^{-12} - 10^{-9} G$, assuming 
an unstructured field along the propagation path.

\paragraph*{Results.} 

\begin{figure}
\begin{center}
\leavevmode\epsfxsize=3.5in\epsfbox{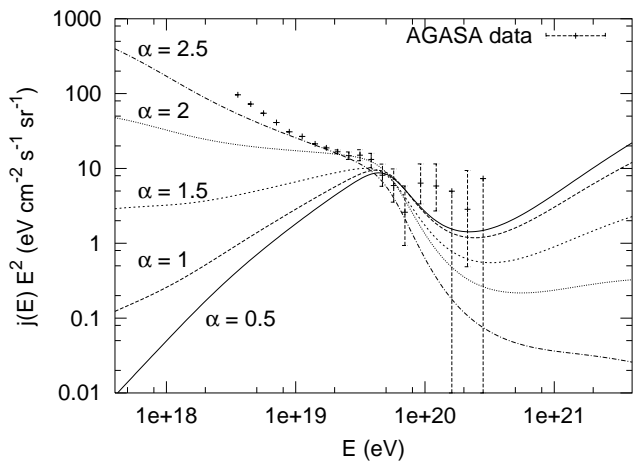}
\end{center}
\caption{Proton spectra for various values of the power law index $\alpha$.}
\label{fig:Cont_prot_alf}
\end{figure}

Astrophysical sources imply acceleration mechanism of the UHECR production,
therefore protons always exist as primaries. We study their propagation first.
We assume power law injection spectra, $J \propto E^{-\alpha}$.
To start with, we study the dependence of the observed spectra
on the value of $\alpha$ assuming homogeneous distribution of sources, no
evolution in comoving volume, and we place no restrictions on the
distance to the nearest source. Resulting spectra are shown in 
Fig~\ref{fig:Cont_prot_alf}.
%%%%%%%%%%%%%%%%%%%%%%%%%%%%%%%%%%%%%%%%%%%%%%%%%%%%%%%%%%%%
%%%%%%%%%%%%%%%%%%%%%%%%%%%%%%%%%%%%%%%%%%%%%%%%%%%%%%%%%%%%

The GZK cut-off is clearly seen in all cases, but its impact is different depending
on $\alpha$. ``Hard'' injection spectra, $\alpha \lsim 1.5$, can be nearly 
reconciled with the data provided some other component of cosmic rays
(Galactic) 
exists at $E \lsim 10^{19}$ eV. Note that injection spectra arising 
in the Z-burst model
can be roughly approximated by $\alpha \lsim 1$ while those arising 
in the decaying
topological defects model can be  approximated by $\alpha \sim 1.5$.
Astrophysical
acceleration mechanisms often result in $\alpha \gsim 2$ \cite{AS2}, however,
harder spectra,  $\alpha \lsim 1.5$ are also possible, see e.g. \cite{AS1.5}.

\begin{figure}
\begin{center}
\leavevmode\epsfxsize=3.5in\epsfbox{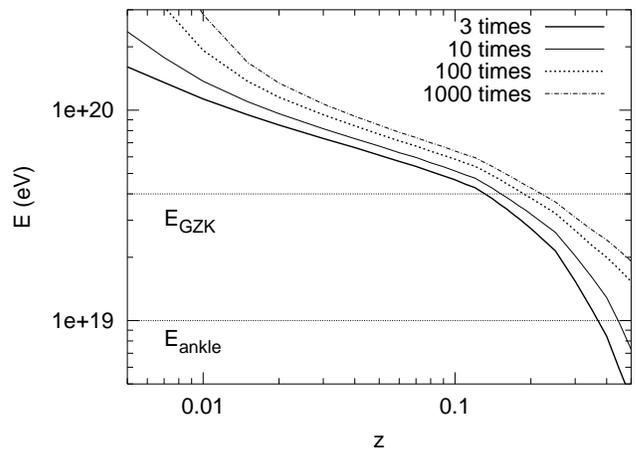}
\end{center}
\caption{Levels of a constant dumping 
of the proton flux as a function of distance traversed.}
\label{fig:Discr_pp1.0_z}
\end{figure}

Different models of UHECR generation can be
discriminated if sources are identified and distances to them are known.
Unfortunately, identity of particular sources is lost in the overall spectrum 
of Fig.~\ref{fig:Cont_prot_alf} and one has to construct the observed spectra
of individual sources as a function of the distance. This procedure
was carried out in Ref. \cite{berez}, however, the wealth of information arising
with this treatment may be prohibitive for presentation in a Letter. 
We represent it in the following way.
First we construct individual spectra as a function of z.
For each given spectra we find the value of energy at which
the number of particles per decade of energy becomes smaller than
the freely propagated particle flux by a given factor. (3, 10, etc.) 
We plot energy thus obtained as a function of z. Results are presented
in Fig.~\ref{fig:Discr_pp1.0_z}. 
We see that curves with an increasing dumping factor converge rapidly
in the range $0.01 \lsim z \lsim 0.5$, therefore, if the
redshift to the source is in this range, Fig. \ref{fig:Discr_pp1.0_z} 
allows to determine maximal proton energies expected from this source. 

The horizontal line at
$E = E_{\rm GZK} \equiv 4 \times 10^{19}$ eV 
corresponds to the formal beginning of the GZK cut-off.
Attenuation length at this energy is $l_a \sim 10^3$ Mpc. This may give a
false impression that protons with $E=E_{\rm GZK}$ reach us from the sources
located at $l=l_a$. Contribution of these protons is negligible as can be
seen from Fig.~\ref{fig:Discr_pp1.0_z}: for $z>0.2$ bulk of the protons
have $E < 4 \times 10^{19}$ eV.

%%%%%%%%%%%%%%%%%%%%%%%%%%%%%%%%%%%%%%%%%%%%%%%%%%%%%%%%%%%%
%%%%%%%%%%%%%%%%%%%%%%%%%%%%%%%%%%%%%%%%%%%%%%%%%%%%%%%%%%%%

We conclude that the contribution of protons to the UHE spectrum 
from distant sources with $z>0.5$ is negligible above AGASA ankle $E>10^{19}$
eV, and it is negligible for sources 
with $z>0.2$ in the highest energy region $E>4 \times 10^{19}$ eV.  

Let us discuss now the propagation and expected spectra of photons.  
Again we consider $\alpha$ as a free parameter. Results are very sensitive
to its value. Interacting with electro-magnetic backgrounds, photons cascade
to low energies which may lead to overproduction of ``soft''
gamma-rays.
Main constraint is given by the EGRET observations in the energy range 
$10^8$ eV - $10^{10}$ eV \cite{EGRET}.
We find that injection power law spectra with indexes $\alpha \ge 2$ cannot 
lead to a sizable 
contribution to the UHECR and obey the EGRET bound simultaneously. 
This is valid even for vanishing EGMF. Therefore, in what follows we consider 
spectra with $\alpha \lsim 2$.
With this restriction the value of EGMF becomes a crucial parameter.

\begin{figure}
\begin{center}
%\rotatebox{-90}
{\centering\leavevmode\epsfxsize=3.4in\epsfbox{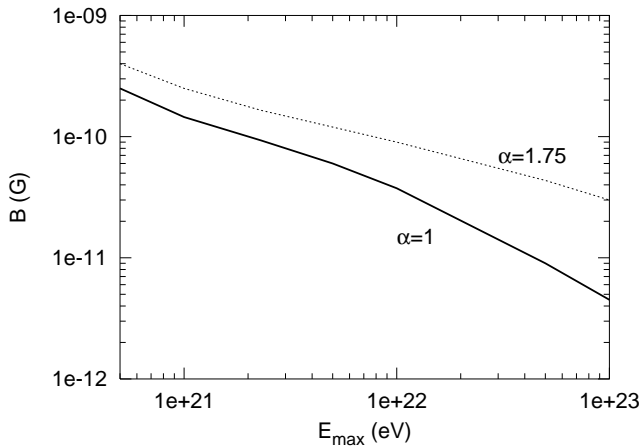}}
\end{center}
\caption{Maximum allowed value of EGMF strength $B$ as a function
of maximal injection energy.}
\label{fig:maxB}
\end{figure}

We have studied the dependence of the resulting photon spectra 
on EGMF and on the maximum energy of 
injected photons for different values of $\alpha$. 
Our first requirement was that the spectra describe highest energy
cosmic ray data well. Our second requirement was that the conflict
with EGRET bound does not appear. For each value of $\alpha$ and
$E_{\rm max}$ this gives maximum possible value of EGMF strength, 
$B$, at which conflict does not appear. This maximum value of B 
does not depend significantly on the spectral shape in the range of $\alpha$ 
we have considered, $1< \alpha < 1.75$, and is
plotted in Fig. \ref{fig:maxB}. Parameter space below line with 
a given value of
$\alpha$ is allowed for this $\alpha$ and leads to resolution of 
the GZK puzzle with photons being primaries.

Note that the dimensionality of the parameter space is actually very large.
In this letter we present only significant dependencies,
while dependence e.g. on cosmological parameters (we assumed 
$H_0 = 70~ {\rm km/s/Mpc}$ and $\Omega_\Lambda = 0.7$) and on the evolution 
of sources (we assumed no evolution having in mind possible correlations
with BL Lacertae) are weak. These less essential dependensies will be discussed
elswere, \cite{large}.
%%%%%%%%%%%%%%%%%%%%%%%%%%%%%%%%%%%%%%%%%%%%%%%%%%%%%%%%%%%%
%%%%%%%%%%%%%%%%%%%%%%%%%%%%%%%%%%%%%%%%%%%%%%%%%%%%%%%%%%%%

In constructing photon spectra which lead to Fig. \ref{fig:maxB}, 
we made no restrictions on the
distance to the nearest source. With such restrictions, i.e. if there 
are no close sources, parameter space is more narrow.
In particular, if there are no sources of UHECR in the GZK volume as 
in the case of BL Lacertae,
one could think that UHE photons cannot reach us without significant 
energy loss.
Indeed, attenuation length of photons is less then 10 Mpc for
energies $10^{19} {\rm eV}<  E < 3 \times 10^{20}$ eV, therefore one can think
that there should be no UHE photon events
with such energies. However, this is not true if the photon injection spectrum
extends to large
energies, $E \gg 10^{21}$ eV. For photons of this energy the attenuation 
length
is as large as several hundred  Mpc. This means that UHE photons originating 
with highest energies at these distances will still be cascading at
energies above the GZK cut-off while approaching us.
As a result they will be continuously recreating secondary 
photons with energies
$10^{19} {\rm eV}<  E < 3 \times 10^{20}$ eV as well. Interestingly, 
we find that 
these secondary photons in this energy range have a power law 
spectrum $1/E^2$ regardless of the
value of $\alpha$ of the initial injection spectrum.
%%%%%%%%%%%%%%%%%%%%%%%%%%%%%%%%%%%%%%%%%%%%%%%%%%%%%%%%%%%%
\begin{figure}
\begin{center}
%\rotatebox{-90}
{\centering\leavevmode\epsfxsize=3.4in\epsfbox{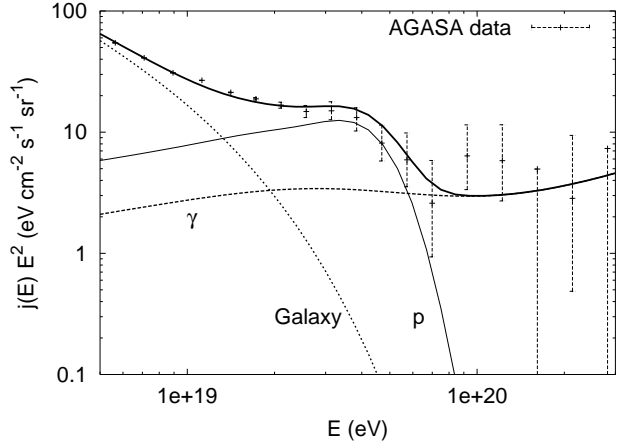}}
\end{center}
\caption{Solid line shows combined contribution of proton and
photon components of UHECR. AGASA data are also shown.}
\label{fig:z=0.03}
\end{figure}
%%%%%%%%%%%%%%%%%%%%%%%%%%%%%%%%%%%%%%%%%%%%%%%%%%%%%%%%%%%%

An example of a resulting UHECR spectrum with crussial assumtion of absence
of sources withing GZK volume is presented in Fig. \ref{fig:z=0.03}.
Here we have assumed that
the closest source in the distribution obeys the condition $ z > 0.03 $ and 
EGMF is small, $B = 10^{-12} G$. We also assume injection spectrum 
$\alpha=1.5$ for both protons and photons.
Resulting proton and photon contributions are shown
separately by thin solid and short-dashed lines respectively. 
We describe the low end of the spectra by independent
Galactic contribution which is modeled by the power law $1/E^{3.16}$
at small energies with an exponential drop at energies around the ankle, $E \sim
10^{19}$ eV. The solid line in Fig. \ref{fig:z=0.03} shows the 
sum of all components.
Photons
starts to dominate total UHECR spectrum with effective power law $1/E^2$ 
at energies $E \sim 5-6 \times 10^{19}$~eV.
Interestingly, this is the value of energy where the clustering
(small angle autocorrelations) in AGASA data set \cite{AGASA} 
becomes most significant \cite{TT_C}.

\begin{table}
\caption{Parameter choices leading to the fit as good 
as in Fig. \ref{fig:z=0.03}.}
\begin{tabular}{cccc}
$z_{\rm min}$&$\alpha$ &  $E_{\rm max}$(eV)   &  $N_\gamma/N_p$ \\ 
\hline
0.03&1.5 &$10^{23}$ & 3 \\
0.03&1.5&$10^{22}$& 17 \\
0.03&1.75&$10^{23}$& 12 \\
0.03&1.75&$10^{22}$& 45 \\
0.1&1.5 &$10^{23}$ & 60 \\
\end{tabular}
\end{table}
The ratio of photons to protons at injection (at given energy) 
which leads to the best fit; in the case of Fig. \ref{fig:z=0.03}
is $N_\gamma/N_p = 3$.
Restrictions
on this parameter and on the maximal energy of injected photons are presented
in Table I for different values of $\alpha$.
Smaller values of $E_{\rm max}< 10^{22}$eV do not work 
(unrealistically large number of photons per proton is required) 
because of the 
rapid decrease of the attenuation length for photons. 
Minimum distance to the closest source at $z = 0.1$ still works 
with the same value of EGMF. However,
this nice picture is destroyed if the EGMF is larger
than a few $\times 10^{-12}$ G.

\paragraph*{Conclusions} 
We have studied spectra of the UHECR assuming primaries are protons and
photons and injection spectrum is a power law $\propto E^{-\alpha}$.
With a homogeneous distribution of sources and 
a hard injection spectra, $\alpha < 1.5$, we find that protons 
can account for the observed flux at highest energies 
producing only a shallow dip around the GZK energy.
Magnitude of the effect is not in strong
disagreement with the data at the level of current statistics.
Presence of (invisible) sources within 
GZK sphere is required, however, if protons are the only primaries. 
Individual sources located at $z > 0.2$ make negligible
contribution into proton component at $E > 4 \times 10^{19}$ eV.
Inclusion of photons makes agreement with the data better. 
In this case even distant sources with $z>0.03$, such as BL Lacertae, can
contribute to observed rays in the energy range
$E > 10^{19}$ eV with the effective power law spectrum $1/E^2$, if
injection spectrum extends up to $E_{\rm max}> 10^{22}$ eV and 
EGMF does not exceeds $10^{-12}$ G .   
Photon component becomes dominant at $E > 5 \times 10^{19} eV$.
In the case when there are sources at $z \lsim 0.1$, the suggested scenario is 
more economical than the Z-burst
model which requires 
acceleration of primaries to even higher energies 
$E_{max}> 10^{23}$ eV.
In addition, the Z-burst model requires extremely large
fluxes of neutrino, while it is enough to have photon flux at the source
to be larger than the proton flux by a factor of only a few.

We conclude that the GZK cut-off can be avoided with photons as primaries
making perfect fit to the data. Parameter space is rather large
if there are no restriction to the distance to the nearest source,
see Fig. \ref{fig:maxB}. 
We cannot rule out photons as primaries even in the
case when production sites are BL Lacertae \cite{TT_BL}, which 
(with known redshifts)
are all outside the GZK volume. To rule it out one needs a
source-by-source study taking into account the concrete configuration
of extragalactic magnetic fields.

\paragraph*{Acknowledgments}
{\tolerance=400 We are grateful to V.S. Berezinsky, G.Sigl and P. Tinyakov 
for valuable comments and discussions.
This work is supported by INTAS grant 99-1065. }

\end{document}